# Emergence of Diffusional Hydrogen Escape in High-$T_c$ Superconducting Calcium Superhydride at Megabar Pressures

Xixi Jia[1,*], Haoran Chen[2,*], Xiaoqiu Ye[3], Jian Lv[4,†], Xitian Zhang[1], Hui Wang[1,‡] and Yansun Yao[5]

[1]*Key Laboratory for Photonic and Electronic Bandgap Materials (Ministry of Education), School of Physics and Electronic Engineering, Harbin Normal University, Harbin 150025, China*

[2]*International Center for Quantum Materials, Peking University, Beijing 100871, China*

[3]*Science and Technology on Surface Physics and Chemistry Laboratory, Jiangyou, 621908, China*

[4]*International Center for Computational Method & Software, College of Physics, Jilin University, Changchun 130012, China*

[5]*Department of Physics and Engineering Physics, University of Saskatchewan, Saskatoon, Saskatchewan S7N 5E2, Canada*

[*]H.C. and X.J. contributed equally to this work.

[†]lvjian@calypso.cn; [‡]wh@fysik.cn

## Abstract

High-pressure metal superhydrides have attracted intense scientific interest due to their remarkable superconducting properties. While superconductivity is known to be sensitive to material composition, compositional variability is often overlooked in metal superhydrides at megabar pressures. Using *ab initio* path-integral simulations, we find that up to 7 % of the hydrogen atoms escape from the 215-kelvin superconducting $CaH_6$ upon decompression from 165 to 123 GPa. This loss of hydrogen leads to an elastic instability at low pressures, an "abnormal" positive pressure dependence of the superconducting $T_c$ and a quantum phase transition from a mixed superconducting-diffusive state to a pure superconducting phase. In this mixed phase, proton diffusivity reaches $10^{-8}$ cm$^2$/s at low temperatures and promotes to $10^{-7}$ cm$^2$/s at room temperature, which elucidates the escape of hydrogen at megabar pressures. Our results are consistent with many "anomalous" experimental observations and highlight the significance of composition effects in high-$T_c$ metal superhydrides.

## Introduction

Hydrogen-rich materials have long been considered important platforms for realizing room-temperature superconductivity [1-7]. This idea is primarily grounded in the Bardeen-Cooper-Schrieffer (BCS) theory [8], which attributes conventional superconductivity to interactions between the electrons at the Fermi level and the lattice vibrations. A straightforward way to maximize this interaction is to increase the Fermi-level density of states while incorporating light constituent elements to achieve high vibrational frequencies, making hydrogen-rich materials a natural choice [9]. Indeed, first-principles calculations frequently reveal a strong correlation between high hydrogen content and high superconducting $T_c$ in hydrogen-rich materials [10], particularly for a class of superconducting interstitial superhydrides (SISHs), where hydrogen atoms fill the interstitials of a host lattice composed of either a single metal element or a multi-component alloy (Fig. S1, [11]). In recent years, experimental studies at megabar pressures have uncovered dozens of SISHs, with those exhibiting high $T_c$ (*e.g.* 150~250 K) generally having hydrogen atomic content exceeding 85 % [12-26].

High-$T_c$ SISHs containing such a high concentration of hydrogen atoms render that their experimental synthesis requires extremely high pressures. Despite the careful design of chemical precursors and synthesis conditions, precise experimental control over the chemical composition of the superconducting sample remains a significant challenge, often leading to deviations from the ideal stoichiometry [12,13]. This raises the question of how to quantify the nonstoichiometric defect, δ. Synchrotron X-ray diffraction (XRD) measurements can estimate the value of δ by analyzing volumetric differences between hydrides and constituent elements, but these measurements are typically limited to specific pressure points, *e.g.* $LaH_{10-\delta}$

with $\delta \approx 0.4$ at 150 GPa [12]. Theoretically, first-principles studies offer a feasible approach to solving the δ-value problem, as early calculations have successfully guided the experimental discovery of nearly all binary high-$T_c$ SISHs [9]. However, for these calculations to accurately reflect experiments, they must take into account the physical effects that significantly impact the experimental results, such as those associated with proton dynamics and defects [27,47-49]. It is therefore necessary to move beyond harmonic approximations and idealized crystal models commonly used in early theoretical work. Encouragingly, recent *ab initio* calculations have quantitatively reproduced the δ value of $LaH_{10-\delta}$ (theoretically 0.46 [27]; experimentally 0.4 [12]) by accounting for the effects of thermal and quantum-activated proton dynamics through path-integral simulations [29,30] on perfect and defective structural models [27].

The sodalite-like $CaH_6$, a representative *bcc*-type SISH, was first predicted in 2012 [3] and successfully synthesized in 2022 [20,21]. Compared to *hcp*- and *fcc*-type SISHs, $CaH_6$ has a simpler crystal structure and valence configuration, *i.e.* hydrogen atoms occupy symmetry-equivalent tetrahedral interstices (*T*-sites, Wyckoff position 12*d*) and no *f* orbitals are involved. Despite the apparent simplicity, the experimentally measured $T_c$ of $CaH_6$ samples synthesized at 170 GPa decreases upon decompression [20], similar to that of *hcp*-$YH_9$ [15] and *fcc*-$LaH_{9.6}$ [12], but opposite to the increasing trend predicted for the ideal $CaH_6$ crystal [3,28]. This discrepancy suggests that previous calculations have overlooked some important features of experimental samples, which may include thermal and quantum-activated nuclear motion, as well as nonstoichiometric defects. Here, by treating them appropriately in our simulations, we find that the hydrogen content variability is crucial in determining the structure, dynamics, and superconductivity of the calcium superhydride, even at megabar pressures.

**Results and Discussion**

Experimental XRD studies determined the pressure-volume (*p*-*V*) relation at room temperature for three superconducting samples of calcium superhydride within the pressure range of 123-190 GPa (Fig. 1(a)) [20]. These measurements naturally include contributions from dynamic effects, which require reciprocal theoretical treatments. Therefore, we simulate the theoretical *p*-*V* relation at 300 K by path-integral molecular dynamics (PIMD) using an isothermal isobaric ensemble [30]. To account for defect effects, we extend our simulations to $CaH_{6-\delta}$ with δ values of 0.26 and 0.5, in addition to pristine $CaH_6$. As shown in Fig. 1(a), comparisons between theoretical and experimental data suggest that defects are rather common in the experimental samples, forming not only during synthesis (*e.g.*, sample 1, with δ averaging about 0.28), but also during decompression (*e.g.* samples 2 and 3, with δ reaching about 0.4 at 120 GPa). The former is a common characteristic of SISHs, attributed to the challenges of controlling reactions at high pressure and temperature [20]. In contrast, the latter is a phenomenon not yet reported in SISHs, indicating that decreasing pressure induce hydrogen escape from the samples.

The H-escape model should satisfy several constraints established by previous experimental studies, *e.g.* those of the XRD patterns and the phase boundary. We start validation with the former, choosing $CaH_6$ and $CaH_{5.72}$ at 160 and 140 GPa as examples after a detailed compositional analysis of the experimental samples (Fig. S2). As shown in Fig. 1(b), $CaH_6$ is in better agreement with the experiments (on sample 3 [20]) than $CaH_{5.72}$ at 160 GPa in terms of the (110) peak position, whereas at 140 GPa, the trend is reversed. The scenario is reminiscent of interstitial hydrides of *bcc* metals (*e.g.* $VH_x$ with $1 < x < 2$) [50], where XRD studies demonstrate the stability of the *bcc* metal sublattice across a wide compositional range. Regarding the phase boundary, XRD experiments show that the *bcc*-Ca sublattice starts to lose stability at about 130 GPa during decompression. We explain this mechanically by analyzing the elasticity of $CaH_6$ and $CaH_{6-\delta}$ at 300 K and comparable pressures (*e.g.* 140-120 GPa). Specifically, we simulate three elastic moduli: $M = C_{11} + 2C_{12} + p$, $M' = C_{11} - C_{12} - 2p$ and $M'' = C_{44} - p$, where $C_{11}$, $C_{12}$ and $C_{44}$ are elastic constants and $p$ is pressure. Negative values of these moduli indicate elastic instability, according to the generalized Born's criteria for cubic crystals under external pressure[36]. As shown in Fig. 1(c), all three moduli remain positive for $CaH_6$, whereas the shear modulus M'' becomes negative at 125 GPa in the $CaH_{6-\delta}$ model as $\delta$ increases during decompression (Table S1). This provides the first example of H-escape-driven elastic instability in SISHs.

In extremely compressed $CaH_6$ (*e.g.* at megabar pressures), hydrogen atoms are squeezed into adjacent interstitial sites and strongly repel each other. The appearance of H defects inevitably induces structural relaxation, which may lead to interstitial proton migration. Indeed, proton hopping was consistently observed in the PIMD simulations of the *p-V* relations and elastic constants of $CaH_{6-\delta}$ (Fig. S3). Taking $CaH_{5.72}$ as an example, we further investigated the diffusivity using ring polymer molecular dynamics (RPMD) in a microcanonical ensemble [31,32]. As shown in Fig. 2(a), the mean square displacements (MSD, $\langle u^2 \rangle$) suggest significant proton diffusion at 300 K, with the diffusion coefficient $D$ reaching $3.8\times10^{-7}$, $2.3\times10^{-7}$ and $0.8\times10^{-7}$ $cm^2/s$ at 140, 160 and 180 GPa, respectively. The $D$ value increases upon decompression and gradually approaches the $10^{-6}$ $cm^2/s$ measured in $LaH_{10+\delta}$ at 160 GPa and room temperature [51]. In $LaH_{10+\delta}$, diffusion induces gradual dehydrogenation of the samples over time, accompanied by a decrease in pressure. Similarly, for $CaH_{6-\delta}$, the large $D$ values suggest that diffusion is the driving force behind hydrogen escape during decompression.

For interstitial hydrides, the variation of diffusivity with temperature is generally more pronounced than with pressure, so it is of great interest to study the temperature effects on proton dynamics in $CaH_{5.72}$, a typical interstitial superhydride. As can be seen from the MSD curves in Fig. 2(a), lowering the temperature from 300 K to 100 K eventually suppresses proton hopping. At 200 K, proton hopping still occurs occasionally alongside structural relaxations, but it appears to be a localized dynamic behavior rather than long-range diffusion. These findings suggest that, at a constant defect concentration, the diffusivity in $CaH_{6-\delta}$ depends monotonically on both temperature and pressure with opposite trends: increasing with temperature but decreasing with pressure.

Low temperature electrical experiments on calcium superhydride samples at different pressures involve room-temperature decompression processes, during which defect concentration may change due to hydrogen escape [20]. As a result, proton dynamics at the experimental superconducting phase transitions are expected to be more complex than in $CaH_{5.72}$ at comparable thermodynamic conditions, as the $\delta$ becomes an additional variable influencing the effects of temperature and pressure on diffusion. To explore this, we investigate proton dynamics in a H-escape $CaH_{6-\delta}$ model where $\delta$ gradually increases from 0.05 to 0.28 during decompression from 160 to 140 GPa. Simulations are carried out at selected temperatures ($T_s$) ranging from 195 to 135 K. As shown in Fig. 2(b), the diffusion is most significant in $CaH_{5.89}$ at 180 K and 155 GPa, rather than in $CaH_{5.94}$ at 195 K and 160 GPa. The $D$ value of the latter ($6.3\times10^{-8}$ $cm^2/s$) is half that of the former ($1.2\times10^{-7}$ $cm^2/s$), which we attribute to proton 'traffic jams' caused by limited vacancies, since the equivalent vacancy diffusivity $D'$ is nearly identical in both compounds.

At 155-140 GPa, temperature effects dominate the proton dynamics in $CaH_{6-\delta}$ ($0.28 \leq \delta \leq 0.11$): neither increasing $\delta$ nor decreasing $p$ prevents the suppression of proton diffusion at low temperatures (*e.g.* 135 K), as shown by the MSD curves and the proton density distributions in Fig. 2(b)-(d). Given that $T_s$ overlaps with the experimental $T_c$ within the 160-140 GPa range [20], the results indicate that hydrogen escape may trigger a novel quantum phase transition in $CaH_{6-\delta}$ during decompression: from a mixed phase of superconductivity and diffusion to a purely superconducting phase.

The sharp decline in experimental $T_c$ observed during decompression from 160 to 140 GPa (Ref. [20]) challenges stoichiometric crystal models of calcium superhydrides, which generally predict a negative pressure dependence of $T_c$ [3,52]. This theoretical trend can be understood qualitatively from the analytical McMillan formula [44,53], where $T_c$ explicitly depends on characteristic parameters of both electrons and phonons. Since the electronic structure is weakly dependent on pressure, phonons dominate the $T_c$ evolution in these crystal models. In $CaH_6$, for example, phonon softening associated with volume expansion enhances the $T_c$ (Table S2). However, the behavior changes dramatically in the H-escape $CaH_{6-\delta}$ model, where δ and volume increase synergistically, fundamentally altering the mechanism. The Fermi-level electronic density of states $N(\epsilon_F)$ decreases with increasing δ (Fig. 3(a)), while the nonmonotonically varying diffusion rate disrupts the trend of continuous phonon softening seen with volume expansion (Fig. 3(b)). In particular, the volume expansion in $CaH_{6-\delta}$ (2%) is only half that of $CaH_6$ (4%), corresponding to a 50 % reduction in pressure variation, thus further strengthening the electron contributions while weakening the phonon contributions to the $T_c$ evolution. Obviously, these changes may lead to a positive pressure dependence of $T_c$ in $CaH_{6-\delta}$, in contrast to the negative dependence predicted for stoichiometric calcium superhydrides.

Quantitatively determining $T_c$ for $CaH_{6-\delta}$ is challenging due to the quantum fluctuations and hydrogen diffusion within its dense H network, which inevitably induce phonon anharmonicity, preventing the application of conventional perturbation theories of superconductivity [48]. To overcome this, we determine the $T_c$ using the recently developed stochastic path-integral approach, which reformulates the Eliashberg equations in a non-perturbative framework, capturing both anharmonic vibrations and diffusion effects under path-integral schemes [39-42]. We first calculate the $T_c$ from the RPMD trajectories generated in the diffusivity studies. As shown in Fig. 3(c), the least defective $CaH_{5.94}$ exhibits the highest $T_c$ (201 K) among the $CaH_{6-\delta}$ phases at 160 GPa, comparable to the 196 K $T_c$ of $CaH_6$ calculated at the same pressure by another nonperturbative method (SSCHA) that accounts for quantum fluctuations [54]. The $T_c$ drops rapidly during decompression, reaching 155 K at 140 GPa. The theoretical pressure variation rate of ~2.3 K/GPa is comparable to the experimental result of 1.9 K/GPa. Changing the simulation method from RPMD to PIMD (at 300 K) slightly increases $T_c$ but does not alter the overall trend. Additional PIMD simulations at 300 K show a weak pressure dependence of $T_c$ in $CaH_6$ at 160-140 GPa, which is inconsistent with the experimental observations. These results establish δ variations in the $CaH_{6-\delta}$ phase as a key factor behind the sharp decrease in $T_c$ observed in the depressurization experiments[20], and theoretically confirm the intriguing quantum phase transition in $CaH_{6-\delta}$ predicted above.

**Summary**

In summary, the present work illustrates the compositional variability and its significant impact on the structural and superconducting properties of high-$T_c$ calcium superhydride at megabar pressures. At room temperature, the presence of non-stoichiometric defects leads to a 'superionic-like' ground state in the calcium superhydride, where thermally and quantum-activated proton diffusion drives hydrogen escape during decompression. This process explains the low-pressure structural instability and sharp decrease in $T_c$ observed in previous experiments. At low temperatures, hydrogen escape induces a novel quantum phase transition during decompression, shifting from a mixed phase of superconductivity and diffusion to a purely superconducting phase. The results highlight the significant "composition effects" in calcium superhydrides and emphasize the need to account for them in future modelling of other high-pressure metal superhydrides--not only to reconcile 'unexpected' experimental observations, but also to accelerate the search for room-temperature superconductivity in this class of quantum materials.

**Acknowledgements** H.W. is thankful to G.T. Liu, H.B. Wang and M. Shiga for valuable discussions, and to L. Ma for the experimental EoS and XRD data. The project is supported by the National Natural Science Foundation of China (Grant No. 12474223 and 11974134), the Team program of the Natural Science Foundation of Heilongjiang Province, China (No.TD2021E005), and the Natural Sciences and Engineering Research Council of Canada.

## Figure Captions

**Figure 1**. (a) The EoSs simulated at 300 K for $CaH_{5.5}$, $CaH_{5.74}$ and $CaH_6$ compared with the experimental data points measured at room temperature by Ma *et. al.* [20]. The insert structure is $CaH_6$, with Ca and H atoms represented by large and small spheres, respectively. (b) The simulated (110) peak position of $CaH_{5.72}$ and $CaH_6$ compared with the experimental XRD patterns from Ma *et. al.* [20], where the green asterisks indicate reflections of $CaH_4$ and the pink asterisk indicate an unidentified reflection. (c) Pressure dependence of the elastic moduli M, M' and M'' in $CaH_{6-\delta}$ and $CaH_6$ derived from PIMD simulations at 300 K.

**Figure 2**. (a) The $\langle u^2 \rangle$ of $CaH_{5.72}$ at pressures of 140-180 GPa and temperatures of 100-300 K. (b) The $\langle u^2 \rangle$ of $CaH_{6-\delta}$ at pressures of 140-160 GPa and various temperatures. (c) Proton density distributions of $CaH_{5.89}$ at 155 GPa and 180 K. (d) Proton density distributions of $CaH_{5.72}$ at 140 GPa and 135 K.

**Figure 3**. (a) The $N(\epsilon_F)$ as a function of $\epsilon_F$ for $CaH_{6-\delta}$ at pressures of 140-160 GPa and temperatures of 135-195 K, with error bar indicating the standard deviations. (b) The vibrational density of states $f(\omega)$ of $CaH_{5.72}$ (140 GPa, 135 K), $CaH_{5.83}$ (150 GPa, 165 K) and $CaH_{5.94}$ (160 GPa, 195 K) derived by Fourier transform of the velocity autocorrelation functions. The logarithmic average of the vibrational frequency $\omega_{\log}$ is also given for comparison. (c) Theoretical $T_c$ of $CaH_{6-\delta}$ and $CaH_6$ calculated by a non-perturbative stochastic path-integral approach compared with experimental data points from Ma *et. al.* [20] and theoretical $T_c$ of $CaH_6$ from Hou *et. al.* [54].

**Figure 1**

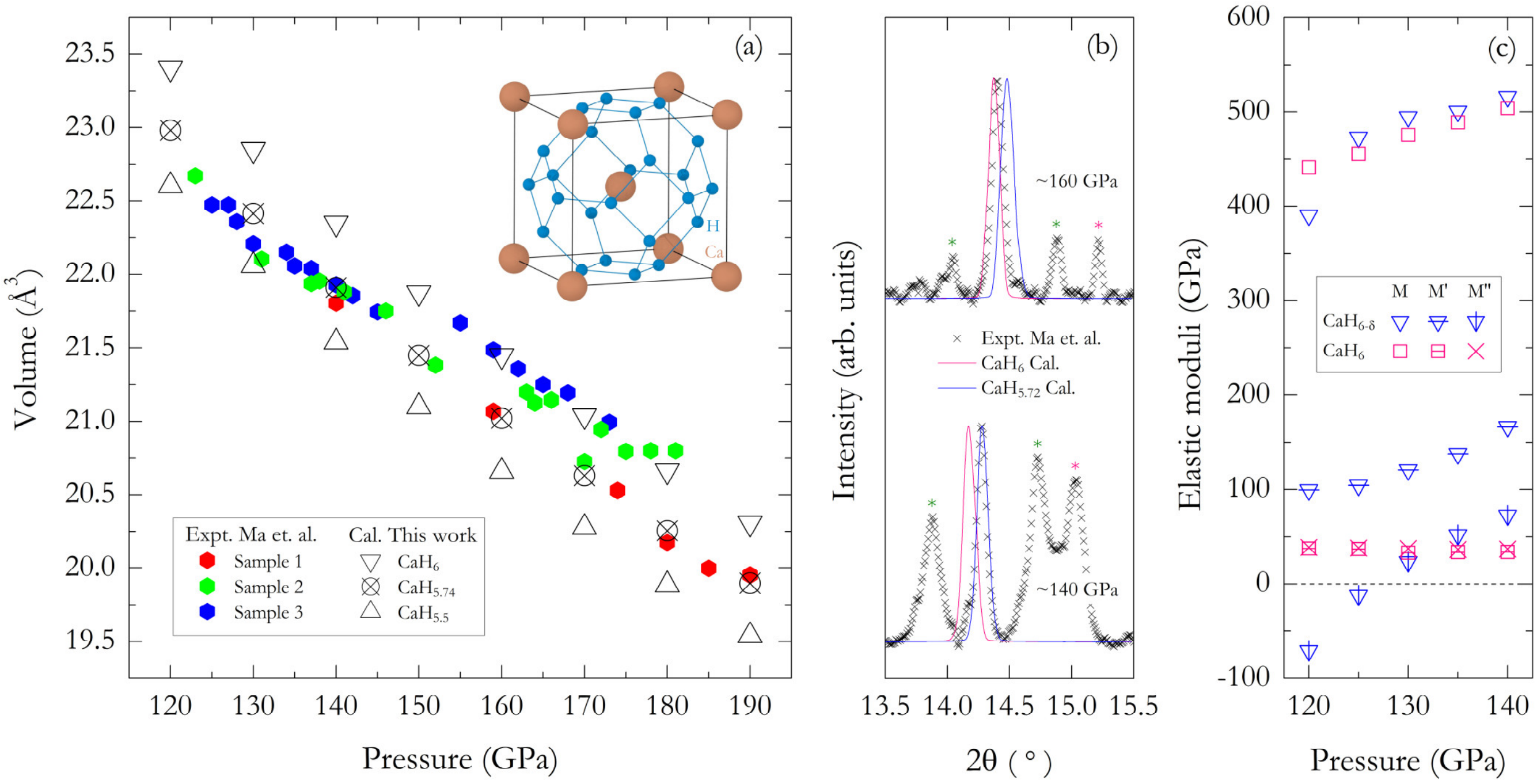

**Figure 1**. (a) The EoSs simulated at 300 K for $CaH_{5.5}$, $CaH_{5.74}$ and $CaH_6$ compared with the experimental data points measured at room temperature by Ma *et. al.* [20]. The insert structure is $CaH_6$, with Ca and H atoms represented by large and small spheres, respectively. (b) The simulated (110) peak position of $CaH_{5.72}$ and $CaH_6$ compared with the experimental XRD patterns from Ma *et. al.* [20], where the green asterisks indicate reflections of $CaH_4$ and the pink asterisk indicate an unidentified reflection. (c) Pressure dependence of the elastic moduli M, M' and M'' in $CaH_{6-\delta}$ and $CaH_6$ derived from PIMD simulations at 300 K.

**Figure 2**

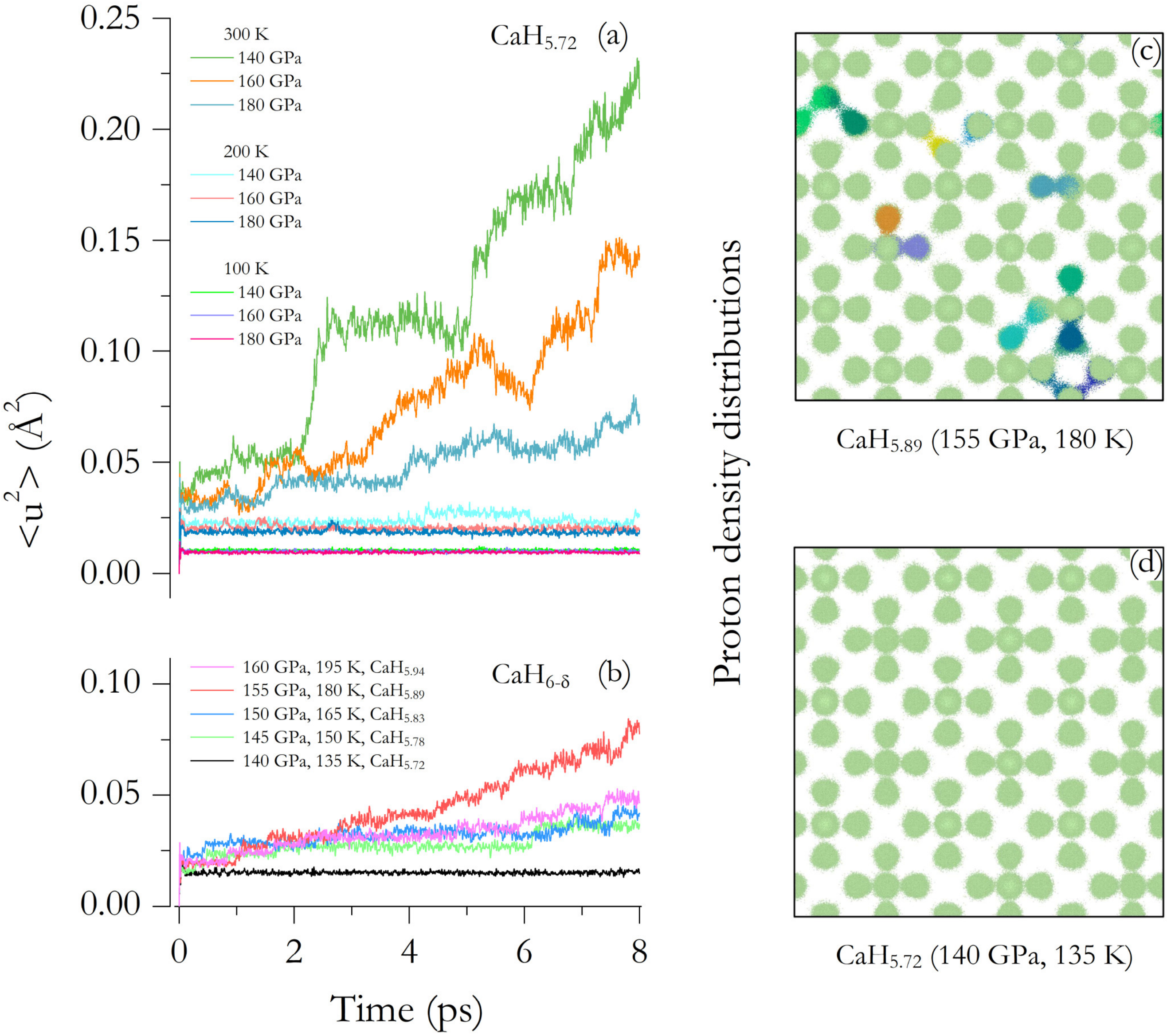

**Figure 2**. (a) The $\langle u^2 \rangle$ of $CaH_{5.72}$ at pressures of 140-180 GPa and temperatures of 100-300 K. (b) The $\langle u^2 \rangle$ of $CaH_{6-\delta}$ at pressures of 140-160 GPa and various temperatures. (c) Proton density distributions of $CaH_{5.89}$ at 155 GPa and 180 K. (d) Proton density distributions of $CaH_{5.72}$ at 140 GPa and 135 K.

**Figure 3**

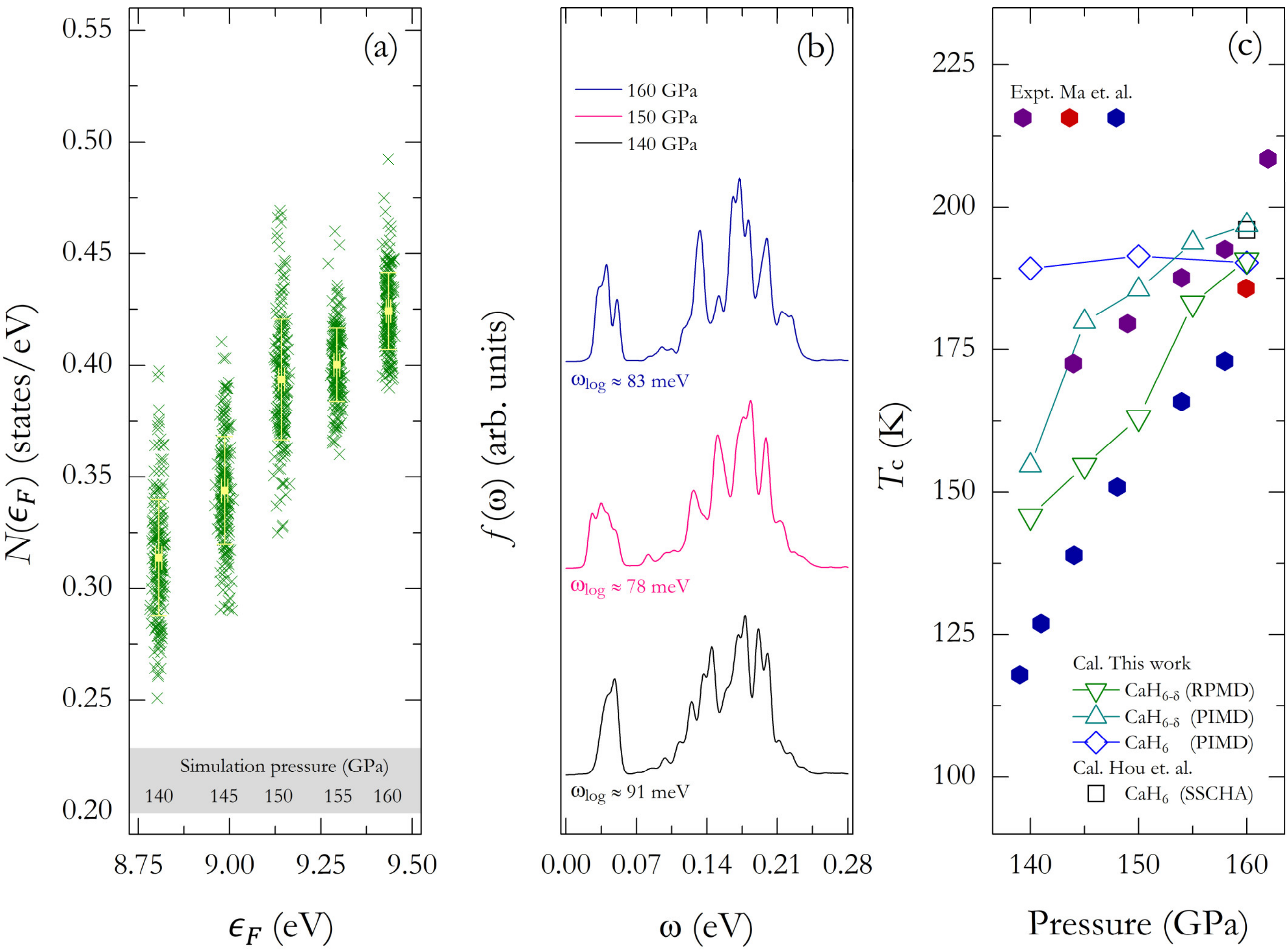

**Figure 3**. (a) The $N(\epsilon_F)$ as a function of $\epsilon_F$ for $CaH_{6-\delta}$ at pressures of 140-160 GPa and temperatures of 135-195 K, with error bar indicating the standard deviations. (b) The vibrational density of states $f(\omega)$ of $CaH_{5.72}$ (140 GPa, 135 K), $CaH_{5.83}$ (150 GPa, 165 K) and $CaH_{5.94}$ (160 GPa, 195 K) derived by Fourier transform of the velocity autocorrelation functions. The logarithmic average of the vibrational frequency $\omega_{\log}$ is also given for comparison. (c) Theoretical $T_c$ of $CaH_{6-\delta}$ and $CaH_6$ calculated by a non-perturbative stochastic path-integral approach compared with experimental data points from Ma *et. al.* [20] and theoretical $T_c$ of $CaH_6$ from Hou *et. al.* [54].

## References

[1] N. W. Ashcroft, Phys. Rev. Lett. **92**, 187002 (2004).

[2] E. Zurek, R. Hoffmann, N. W. Ashcroft, A. R. Oganov, and A. O. Lyakhov, Proc. Natl. Acad. Sci. **106**, 17640 (2009).

[3] H. Wang, J. S. Tse, K. Tanaka, T. Iitaka, and Y. Ma, Proc. Natl. Acad. Sci. **109**, 6463 (2012).

[4] F. Peng, Y. Sun, C. J. Pickard, R. J. Needs, Q. Wu, and Y. Ma, Phys. Rev. Lett. **119**, 107001 (2017).

[5] K. Dolui, L. J. Conway, C. Heil, T. A. Strobel, R. P. Prasankumar, and C. J. Pickard, Phys. Rev. Lett. **132**, 166001 (2024).

[6] G. S. Boebinger *et al.*, Nat. Rev. Phys. **7**, 2 (2025).

[7] D. V. Semenok, F. Bärtl, D. Zhou, T. Helm, S. Luther, H. Kühne, J. Wosnitza, I. A. Troyan, and V. V. Struzhkin, arxiv.org/abs/2412.11727 (2025).

[8] J. Bardeen, L. N. Cooper, and J. R. Schrieffer, Phys. Rev. **108**, 1175 (1957).

[9] J. A. Flores-Livas, L. Boeri, A. Sanna, G. Profeta, R. Arita, and M. Eremets, Phys. Rep. **856**, 1 (2020).

[10] C. J. Pickard, I. Errea, and M. I. Eremets, Ann. Rev. Condens. Matt. Phys. **11**, 57 (2020).

[11] See Supplemental Material at xxx for supplemental figures and computational details, which includes Refs. [12-46].

[12] A. P. Drozdov *et al.*, Nature **569**, 528 (2019).

[13] M. Somayazulu, M. Ahart, A. K. Mishra, Z. M. Geballe, M. Baldini, Y. Meng, V. V. Struzhkin, and R. J. Hemley, Phys. Rev. Lett. **122**, 027001 (2019).

[14] D. V. Semenok *et al.*, Mater. Today **33**, 36 (2020).

[15] P. Kong *et al.*, Nat. Commun. **12**, 5075 (2021).

[16] D. V. Semenok *et al.*, Mater. Today **48**, 18 (2021).

[17] I. A. Troyan *et al.*, Adv. Mater. **33**, 2006832 (2021).

[18] J. Bi *et al.*, Nat. Commun. **13**, 5952 (2022).

[19] J. Bi *et al.*, Mater. Today. Phys. **28**, 100840 (2022).

[20] L. Ma *et al.*, Phys. Rev. Lett. **128**, 167001 (2022).

[21] Z. Li *et al.*, Nat. Commun. **13**, 2863 (2022).

[22] W. Chen, X. Huang, D. V. Semenok, S. Chen, D. Zhou, K. Zhang, A. R. Oganov, and T. Cui, Nat. Commun. **14**, 2660 (2023).

[23] S. Chen, Y. Wang, F. Bai, X. Wu, X. Wu, A. Pakhomova, J. Guo, X. Huang, and T. Cui, J. Am. Chem. Soc. **146**, 14105 (2024).

[24] S. Chen, J. Guo, Y. Wang, X. Wu, W. Chen, X. Huang, and T. Cui, Phys. Rev. B **109**, 224510 (2024).

[25] S. Chen, Y. Qian, X. Huang, W. Chen, J. Guo, K. Zhang, J. Zhang, H. Yuan, and T. Cui, Natl. Sci. Rev. **11**, nwad107 (2024).

[26] G. Huang *et al.*, J. Phys.: Condens. Matter **36**, 075702 (2024).

[27] H. Wang, P. T. Salzbrenner, I. Errea, F. Peng, Z. Lu, H. Liu, L. Zhu, C. J. Pickard, and Y. Yao, Nat. Commun. **14**, 1674 (2023).

[28] Y. Quan, S. S. Ghosh, and W. E. Pickett, Phys. Rev. B **100**, 184505 (2019).

[29] R. P. Feynman, *Statistical Mechanics: A Set of Lectures* (Hachette, UK, 1998).

[30] M. Shiga, M. Tachikawa, and S. Miura, J. Chem. Phys. **115**, 9149 (2001).

[31] M. Shiga and A. Nakayama, Chem. Phys. Lett. **451**, 175 (2008).

[32] I. R. Craig and D. E. Manolopoulos, J. Chem. Phys. **121**, 3368 (2004).

[33] G. Kresse and J. Furthmüller, Phys. Rev. B **54**, 11169 (1996).

[34] J. P. Perdew, K. Burke, and M. Ernzerhof, Phys. Rev. Lett. **77**, 3865 (1996).

[35] P. E. Blöchl, Phys. Rev. B **50**, 17953 (1994).

[36] J. Wang, J. Li, S. Yip, D. Wolf, and S. Phillpot, Physica A **240**, 396 (1997).

[37] Y. Li, L. Vočadlo, and J. P. Brodholt, Comput. Phys. Commun. **273**, 108280 (2022).

[38] P. E. Blöchl, O. Jepsen, and O. K. Andersen, Phys. Rev. B **49**, 16223 (1994).

[39] H. Liu, Y. Yuan, D. Liu, X. Li, and J. Shi, Phys. Rev. Res. **2**, 013340 (2020).

[40] H. Chen, X. Zhang, X. Li, and J. Shi, Phys. Rev. B **104**, 184516 (2021).

[41] H. Chen and J. Shi, Phys. Rev. B **106**, 184501 (2022).

[42] H. Chen and J. Shi, Phys. Rev. B **109**, L140505 (2024).

[43] X. Zhang, H. Chen, E. Wang, J. Shi, and X. Li, Phys. Rev. B **105**, 155148 (2022).

[44] P. B. Allen and R. C. Dynes, Phys. Rev. B **12**, 905 (1975).

[45] P. Morel and P. W. Anderson, Phys. Rev. **125**, 1263 (1962).

[46] G. D. Mahan, *Many-Particle Physics* (Kluwer Academic, Dordrecht, 2000).

[47] I. Errea *et al.*, Nature **578**, 66 (2020).

[48] I. Errea, J. Phys.: Condens. Matter **34**, 231501 (2022).

[49] H. Chen, H. Wang, and J. Shi, arxiv.org/abs/2409.09836 (2024).

[50] E. Akiba and M. Okada, MRS Bull. **27**, 699 (2002).

[51] Y. S. Zhou *et al.*, Nat. Commun. **16**, 1135 (2025).

[52] D. An, D. Duan, Z. Zhang, Q. Jiang, T. Ma, Z. Huo, H. Song, and T. Cui, Phys. Rev. B **110**, 054505 (2024).

[53] W. L. McMillan, Phys. Rev. **167**, 331 (1968).

[54] P. Hou, Z. Huo, and D. Duan, J. Phys. Chem. C **127**, 23980 (2023).